\documentclass[%
twocolumn,
superscriptaddress,
amsmath,amssymb,
aps,
]{revtex4-1}

\usepackage{graphicx}
\usepackage{dcolumn}
\usepackage{bm}
\usepackage{xcolor}
\usepackage[normalem]{ulem}
\usepackage[mathlines]{lineno}

\usepackage{hyperref}
\hypersetup{
 linktocpage = true,
     colorlinks,
    citecolor=blue,
    filecolor=black,
    linkcolor=blue,
    urlcolor=blue
}

\begin{document}

\preprint{APS/123-QED}

\title{Pruning a restricted Boltzmann machine for quantum state reconstruction}

\author{Anna Golubeva}
\affiliation{Department of Physics and Astronomy, University of Waterloo, Ontario, N2L 3G1, Canada}
\affiliation{Perimeter Institute for Theoretical Physics, Waterloo, Ontario N2L 2Y5, Canada}
\affiliation{Department of Physics, Massachusetts Institute of Technology, Cambridge, MA 02139, USA}
\affiliation{The NSF AI Institute for Artificial Intelligence and Fundamental Interactions}
\author{Roger G. Melko}
\affiliation{Department of Physics and Astronomy, University of Waterloo, Ontario, N2L 3G1, Canada}
\affiliation{Perimeter Institute for Theoretical Physics, Waterloo, Ontario N2L 2Y5, Canada}

\date{\today}

\begin{abstract}
Restricted Boltzmann machines (RBMs) have proven to be a powerful tool 
for learning quantum wavefunction representations from qubit projective measurement data. 
Since the number of classical parameters needed to encode a quantum wavefunction 
scales rapidly with the number of qubits, 
the ability to learn efficient representations is of critical importance.
In this paper we study magnitude-based pruning as a way to compress the wavefunction representation in an RBM, 
focusing on RBMs trained on data from the transverse-field Ising model in one dimension. 
We find that pruning can reduce the total number of RBM weights, 
but the threshold at which the reconstruction accuracy starts to degrade 
varies significantly depending on the phase of the model.
In a gapped region of the phase diagram, the RBM admits pruning over half of the weights 
while still accurately reproducing relevant physical observables.
At the quantum critical point however, even a small amount of pruning can lead 
to significant loss of accuracy in the physical properties of the reconstructed quantum state.
Our results highlight the importance of tracking all relevant observables as their sensitivity varies strongly with pruning.
Finally, we find that sparse RBMs are trainable and discuss how a successful sparsity pattern can be created without pruning.
\end{abstract}

\maketitle

\section{Introduction}\label{sec:intro}

{\it Pruning} is a common technique employed in machine learning (ML) 
with the goal of reducing the overall number of parameters in a neural network (NN) 
in order to make it more memory- and compute-efficient.  
The fact that pruning can substantially reduce the number of weights 
without degrading the performance of a NN has been known for decades by the ML community~\citep{LeCun_1990,Reed_1993,Hassibi_1993}.
A more recent explosion of interest in this technique was brought about by deep learning, 
as deep NNs -- particularly the ones that perform best -- 
require enormous amounts of computational resources.
An efficient compression of such deep NN models is of immediate concern, 
particularly in industrial applications~\cite{Han_2015,hinton2015distilling}.
The most successful approach has been {\it iterative magnitude-based pruning post training}, 
which eliminates the smallest weights from a trained NN stepwise 
and allows the remaining parameters to adjust 
by doing some training iterations after each pruning step~\citep{Thimm1995, Strom1997, Han2016}.
This method can substantially reduce the computational cost of inference and
enable deployment of NN-based applications in resource-constrained environments, 
such as mobile devices~\citep{Han_2015, yang2017designing, sze2017efficient}.
However, pruning {\it post training} means that the NN still must initially be trained at its full size.
A variety of recent works propose pruning based on alternative criteria~\citep{SNIP, evci2019rigging, GraSP}, 
but to date there is no competitive approach to pruning {\it before training}~\citep{Frankle_2020}.
Moreover, the principles of why pruning works in general are not understood 
and it is unclear what determines a successful sparsity pattern for a given NN and learning task. 
For a recent review of the current state of pruning research in ML see Ref.~\cite{Blalock_2020}.

As all fields involving enormous amounts of computational resources, physics would benefit from more efficient NN models. 
In particular in quantum many-body physics 
we are ultimately interested in modeling large physical systems, 
and the anticipated power of ML techniques lies in their ability to handle system sizes 
that conventional numerical methods can not. 
Physics, in turn, has the potential to advance 
the theoretical understanding of pruning and sparsity in NNs.
Specifically, physics problems offer a practical test bed for ML methods because their datasets are well characterized. 
For instance, in case of a quantum many-body problem, the knowledge of the Hamiltonian allows us to describe the entire state space, 
to generate input data using numerical simulations, 
and to derive relevant characteristics of the underlying probability distribution.
This opens up more ways to evaluate the learning procedure and the obtained predictions, 
and thus to gain insights into the processes of NN learning and pruning.

In this paper, we investigate pruning based on the problem of 
quantum state reconstruction with restricted Boltzmann machines (RBMs) \cite{Melko_2019_RBMs}. 
The reconstruction of a quantum many-body state from data 
is one of the most important applications in physics 
requiring NNs with as few parameters as possible. 
In the simplest case of a quantum state prepared on a number of qubits, 
the task involves learning the probability distribution underlying 
a set of projective measurements as given by the Born rule. 
This is most immediately accomplished by a {\it generative model}~\cite{GM_survey}, 
whose parameters are trained to maximize the likelihood that 
it accurately represents the data distribution. 
The success of pruning in the discriminative setting opens up the possibility that a similar strategy might work in the generative setting. 

We choose to work with RBMs~\cite{Smolensky_1986} 
that are among the simplest and oldest generative models in ML. 
RBMs have become familiar to condensed matter and quantum information physicists 
due to their foundations in the Ising model of statistical mechanics. 
Despite their relative simplicity, 
RBMs have a high representational capacity; 
combined with efficient training heuristics, this makes RBMs 
compelling candidates for data-driven state reconstructions~\cite{Torlai2016thermo}.
Several works in quantum physics have demonstrated the use of RBMs for both 
data-driven wavefunction reconstruction~\cite{torlai2018tomography,TorlaiMixed} and as a variational ansatz, 
where parameters are optimized with knowledge of the Hamiltonian~\cite{Carleo}.
Furthermore, RBMs have proven particularly effective in reconstructing quantum states
in experimental devices where shot budgets may be low~\cite{Rydberg}.

The full connectivity of the RBM provides an upper bound on the entanglement 
that it can represent in a physical wavefunction~\cite{Deng2017}. 
This upper bound -- corresponding to the ``volume law'' scaling of entanglement entropy -- 
is sufficiently expressive to capture the entanglement behavior 
of any known class of quantum many-body states. 
However, ground states of local Hamiltonians are expected to obey {\it sub}-volume-law scaling \cite{Hastings_2007}, 
which raises the question whether RBMs with sparser connectivity can be found 
that would lead to more efficient reconstruction schemes.

In this paper, we study the effects of pruning an RBM trained on data from a 
prototypical quantum many-body wavefunction -- the groundstate of the 
one-dimensional transverse-field Ising model (TFIM),
\begin{equation}\label{TFIM}
    H = - J\sum_{\langle i j \rangle} \sigma^z_i \sigma^z_j + h \sum_i \sigma^x_i\,. 
\end{equation}
Here, ${\bm \sigma}_i$ is the Pauli operator representing 
a qubit on site $i$ of the one-dimensional lattice 
with open boundary conditions.
As described in detail below, for various system sizes and transverse field strengths $h/J$,
we use the Density Matrix Renormalization Group (DMRG) algorithm~\cite{White_1992_DMRG,White_1993_DMRG,Ferris, ITensor}
to produce samples of the ground state wavefunction in the $\sigma^z$ basis.
Then, we train RBMs using the Qucumber software package~\cite{qucumber}
and subsequently apply pruning, with the goal of determining how 
the accuracy of reconstruction of the RBM wavefunction is affected. 
In addition to the standard loss function, the Kullback–Leibler (KL) divergence, 
we examine the effect of pruning on a number of relevant physical properties, namely  
the fidelity, energy, order parameter and two-point correlation function.

\begin{figure}[ht]
    \includegraphics[width=0.7\columnwidth]{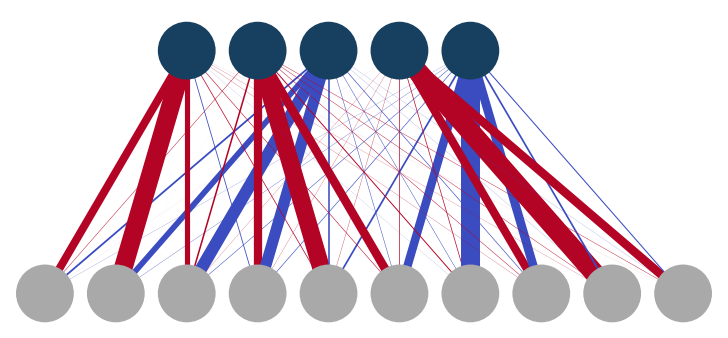}
    \includegraphics[width=0.7\columnwidth]{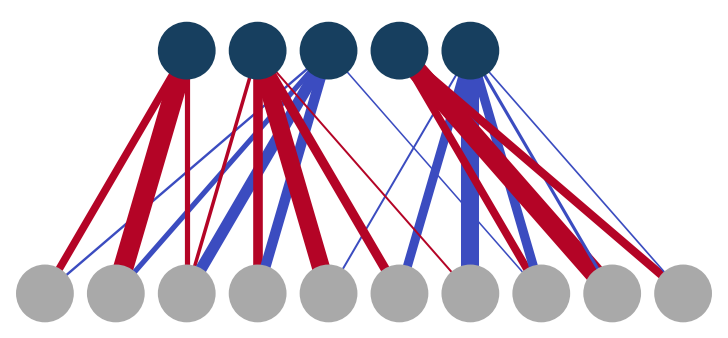}
    \includegraphics[width=0.9\columnwidth]{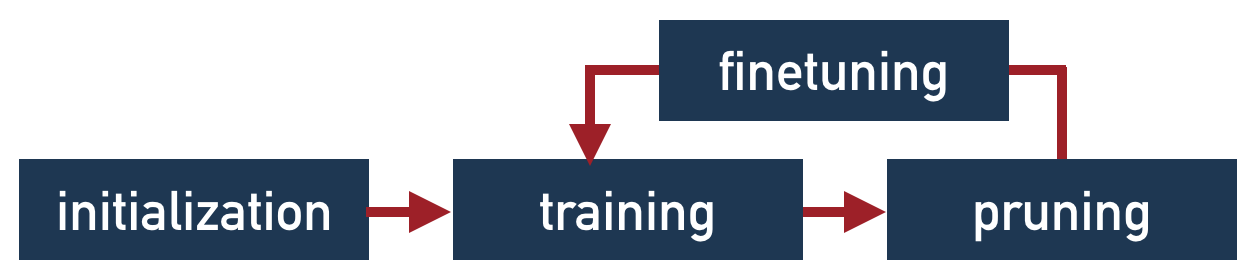}
    \caption{
        \textit{Top:} The graphs of a trained RBM before (upper) and after (lower) pruning. 
        Lines represent weights, line thickness corresponds to weight magnitude 
        and color indicates the sign (red for $+$, blue for $-$).
        \textit{Bottom:} The procedure of training and iterative pruning with finetuning.}
    \label{fig:Fig1}
\end{figure}

\section{Methods and observables}\label{sec:Methods}

When used as a generative model, a standard RBM can be easily trained on 
projective measurements of the groundstate of Eq.~\eqref{TFIM} \cite{torlai2018tomography,Sehayek2019}, 
which is stoquastic in the $\sigma^z$ basis \cite{Bravyi2008}. 
This fact also ensures that the wavefunction can be fully represented by 
the marginal distribution of the RBM according to the Born rule
\begin{equation}
    \psi({{\bm \sigma}}) = \sqrt{p_{\lambda}({\bm \sigma)}}.
\end{equation}

The RBMs that we use in this paper to reconstruct the groundstate of Eq.~\eqref{TFIM} 
are standard and described in detail in~\cite{qucumber}.
The RBM probability distribution corresponding to the wavefunction is encoded in real parameters $\lambda$, 
which include weights and biases.
The weights mediate correlations in the wavefunction 
by connecting the {\it visible} variables (the qubit states) with 
the {\it hidden} variables. 
For a fixed number of visible units $N$, 
the expressivity of the RBM is modified by varying the number of hidden units $N_h$, 
which is usually treated as a hyperparameter and fixed at the beginning of training.
Based on a previous study~\cite{Sehayek2019}, 
we use the ratio $N_h=N/2$ in our numerical calculations below. 

\subsection{Pruning and Sparsity}

The idea of pruning comes naturally when we examine the weights in a trained RBM. 
For instance, Figure~\ref{fig:Fig1} illustrates the graph of 
a fully trained RBM used to reconstruct the groundstate of a 10-qubit TFIM, 
before and after the pruning procedure (described more below). 
It is apparent that after training 
only a small number of weights have large magnitudes and 
all other weights are much smaller in comparison.
This suggests that the small-magnitude weights might be insignificant, 
and could thus be set to zero (``pruned'') without affecting model performance.
As a result of the pruning procedure, the RBM graph becomes more sparsely connected 
and the corresponding RBM still performs adequately according to \textit{some} metrics. 
However, as we show in detail in Section~\ref{sec:results-Dense}, 
pruning impacts various physical qualities in different proportions.

The standard method of pruning and fine-tuning proceeds by eliminating 
the weights of a dense RBM after training is complete.
An interesting question to ask is whether an RBM that is sparse already \emph{at initialization} 
can be trained to achieve the same results as a dense RBM. 
A recent work in ML~\cite{LTH} has shown that 
this is possible for fully-connected feed-forward NNs.
Specifically, if the sparsity pattern obtained by iterative magnitude-based pruning is applied to the NN with all non-pruned weights reset to their initial values, 
the resulting sparse NN can be trained to achieve the same or even better performance as the original dense one. 
A multitude of follow-up works explore this research direction (see the recent review article~\cite{Blalock_2020} for a list of references). 
However, how to find a method to obtain a working sparsity pattern for a NN 
without first training the dense NN remains an open research question. 
Most importantly, it is not yet understood
why NNs are amenable to pruning post training, 
but do not perform as well if they have less parameters from the start. 
A common hypothesis is that NN overparametrization 
-- that is, the superfluous weights -- is required for successful optimization, while 
the final optimized NN model involves only a fraction of all weights.
We investigate the performance of RBMs that are sparse at initialization time in Section~\ref{sec:results-Sparse}.

\subsection{Training metrics and physical estimators}
We train our RBMs using the standard contrastive divergence (CD) algorithm~\cite{Hinton_2002_CD}.
This training procedure is implicitly minimizing the KL divergence
between the target distribution $q$ estimated from the dataset 
and the RBM distribution $p_\lambda$:
\begin{equation}
	D_\text{KL}(q\| p_{\lambda}) = \sum\limits_{{\bm \sigma}} q({\bm \sigma}) \ln \frac{q({\bm \sigma})}{p_{\lambda}({\bm \sigma})}.
\end{equation}
In general, we cannot expect to reach a KL divergence of zero, 
yet any target threshold we set for the KL would be arbitrary, 
because the KL is not calibrated and unbounded from above.
Therefore, in addition to the KL divergence, 
we introduce and track error measures for various physical observables.

The KL divergence measures the discrepancy between 
the RBM-learned and the target distribution.
Similarly, the {\it fidelity} measures the agreement between 
the reconstructed state and the exact wavefunction. For pure states, 
the fidelity is given by 
\begin{equation} \label{eq:F}
	F\left( \psi_\lambda, \Psi \right) = |\langle \psi_\lambda | \Psi \rangle |^2\,.
\end{equation}
For smaller systems (up to $N=18$ qubits), we compute the KL and fidelity explicitly.
However, the computation cost scales exponentially with the system size, 
as it requires summing over the entire state space.
Therefore, 
in order to evaluate the quality of the learned RBM distribution for larger systems, 
we have to rely on other physical observables.

Two characteristic observables for spin systems are the energy $E=\langle H\rangle$ and 
the magnetization along the $z$-axis, defined as
\begin{equation}\label{eq:Magnet}
    m = \frac{1}{N}\sum\limits_i^{N} \sigma_i^z,
\end{equation}
which serves as the order parameter. 
Since the training set comprises simulated configurations of a finite-size system,
it is $\mathbb{Z}_2$ symmetric -- that is, each configuration $v$ and its counterpart $-v$ 
appear in approximately equal proportions. 
Consequently, positive and negative contributions to $\langle m \rangle$ 
would cancel out.
We therefore compute both $\langle m \rangle$ and 
the $\mathbb{Z}_2$-invariant quantity $\langle |m| \rangle$.
As suggested in~\cite{Sehayek2019}, we measure the relative observable error (ROE) 
for the energy and the absolute magnetization, defined as 
\begin{equation}
    \text{ROE} =\max \left|\frac{O_\text{DMRG}-\bar{O}_\text{RBM}}{O_\text{DMRG}}\right|.
\end{equation}
Here, $O_\text{DMRG}$ is the expectation value for a general observable $O$ calculated via DMRG, 
which we consider as exact. 
The RBM estimator is 
$\bar{O}_\text{RBM}$ with a statistical error correction 
calculated as 
$\bar{O}_\text{RBM} = \langle O \rangle_\text{RBM} \pm c \omega/\sqrt{n}$,
where $n$ is the number of samples, 
$\omega$ is the standard deviation, 
and $c=2.576$ is a constant corresponding to $99\%$ confidence interval. 
We shall use the short-form notation ``eROE'' and ``mROE'' 
for energy and magnetization ROE, respectively.

The two-point correlation function is another important observable.
In particular the functional form of its decay with increasing spin distance 
is a distinctive feature of a physical phase.  In the groundstate phase diagram
of the TFIM,
one expects algebraic decay at the quantum critical point (QCP) 
and exponential decay in the gapped phases away from criticality.
We compute 
\begin{equation}
    C(d_{i,j})= 
        \langle \sigma^z_i \sigma^z_j \rangle 
        - \langle \sigma^z_i \rangle \langle \sigma^z_j \rangle, \label{eq:SSCF}
\end{equation}
where $d_{ij}$ is the distance between the sites $i$ and $j$.
To summarize the deviation between the two-point correlation function computed on 
spin configurations sampled from the RBM and the one computed on DMRG training data 
in a single scalar quantity, we use the mean squared error
\begin{equation}
    C_\text{MSE} = \sum \limits_{j=\bar{i}}^{N} C(d_{\bar{i}, j}) \, . \label{eq:SSCF_MSE}
\end{equation}
Here, $\bar{i}=N/2$ is the lattice midpoint and 
the sum extends over all spin distances from 0 to $N/2$, 
as we are working with open boundary conditions.

\section{Results}\label{sec:Results}

\begin{figure*}[ht]
    \includegraphics[width=0.65\textwidth]{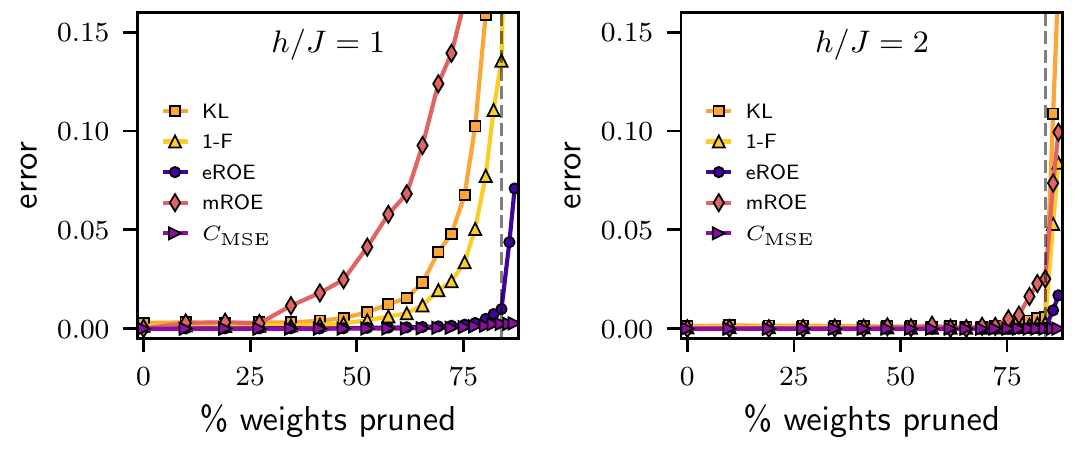}
    \vspace{-2mm}
    \caption{The behavior of various model quality measures as a function of iterative pruning 
    for an RBM of size $18\times 9$, trained at the quantum critical point at $h/J=1$ (\textit{left}), or  
    in the paramagnetic regime at $h/J=2$ (\textit{right}). 
    Grey dashed vertical line at 85\% indicates the point where the RBM graph becomes disconnected.
    }
\label{fig:error_vs_percent_pruned}
\end{figure*}

In this section we present numerical results for our study of RBM pruning. 
The dataset we use for training the RBM was obtained using DMRG~\cite{Sehayek2019,Ferris}. 
It is composed of projective qubit measurements in the $\sigma^z$ basis 
for the groundstate wavefunctions of Eq.~\eqref{TFIM}. 
We focus on two points of the phase diagram: 
the quantum critical point (QCP) at $h/J=1$ 
and inside the paramagnetic (PM) phase at $h/J=2$. 
We train RBMs for a fixed number of epochs, ensuring that all observables have fully equilibrated. 
All expectation values are computed on $10^5$ configurations drawn from a trained RBM. 

\subsection{Pruning Dense RBMs} \label{sec:results-Dense}
We train a dense (fully connected) RBM and
apply iterative magnitude-based pruning and fine-tuning, 
meaning that weights are removed in steps and the model is trained for several more epochs 
after each pruning step (see Fig.~\ref{fig:Fig1}). 
Note that we prune only the weights, not the biases.
Starting from the trained model, we remove 10\% of all non-pruned weights in each step 
-- for details see Appendix, Section~\ref{appx:schedule}.

In the following, we present most of our results for a system of size $N=18$, 
since this is the maximal size for which the computation of 
KL divergence and state fidelity is feasible.
For larger systems, we compute the eROE, mROE and the 2-point correlation functions, 
and find no qualitative difference to the smaller system. 
Furthermore, we examine the correlations between KL divergence and 
the alternate physical error measures of the last section, 
and generally find that all measures are positively correlated with KL. 
However, the correlation is different for each measure, 
and it varies strongly depending on the stage of training or pruning. 
Further details, including more discussion on training, can be found in section~\ref{appx:ROE_KL_correlat} in the Appendix.

Figure~\ref{fig:error_vs_percent_pruned} demonstrates how the various quantities 
that measure model quality evolve in the course of pruning. 
While convergence to the equilibrated values occurs within the first 500 training epochs for all observables (the training part is not shown in the figure), 
their degradation in response to pruning varies and is strongly dependent on the physical phase.
The RBM trained to model the groundstate in the PM regime  
does not show increase in any of the error measures until 
about 75\% of weights have been removed.
In contrast, the model trained at the QCP 
allows us to remove only about 25\% of the weights and further pruning incurs significant damage.

\begin{figure*}[ht]
    \includegraphics[width=0.65\textwidth]{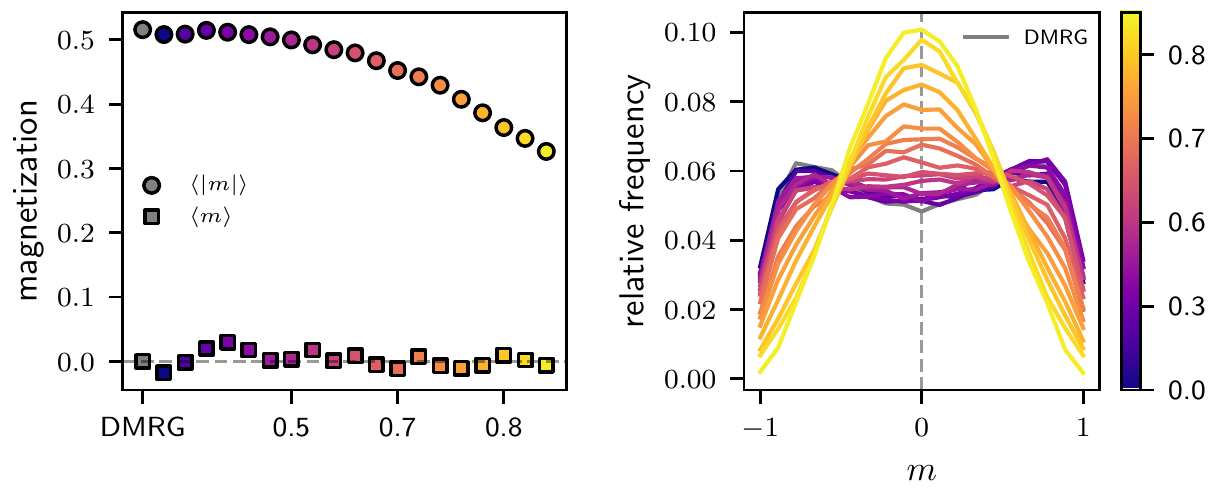}
    \vspace{-3mm}
    \caption{Magnetization expectation values $\langle m \rangle$ and $\langle |m| \rangle$ (\textit{left}), 
    and histogram of the $m$ values for each configuration in a sample of $10^5$ configurations (\textit{right}) at the QCP.
    The numbers on the x-axis in the left plot and in the legend on the right indicate the fraction of weights pruned. 
    The training set, labeled as ``DMRG", serves as reference.
    }
    \label{fig:Magnet}
\end{figure*}

\begin{figure*}[ht]
    \includegraphics[width=0.8\textwidth]{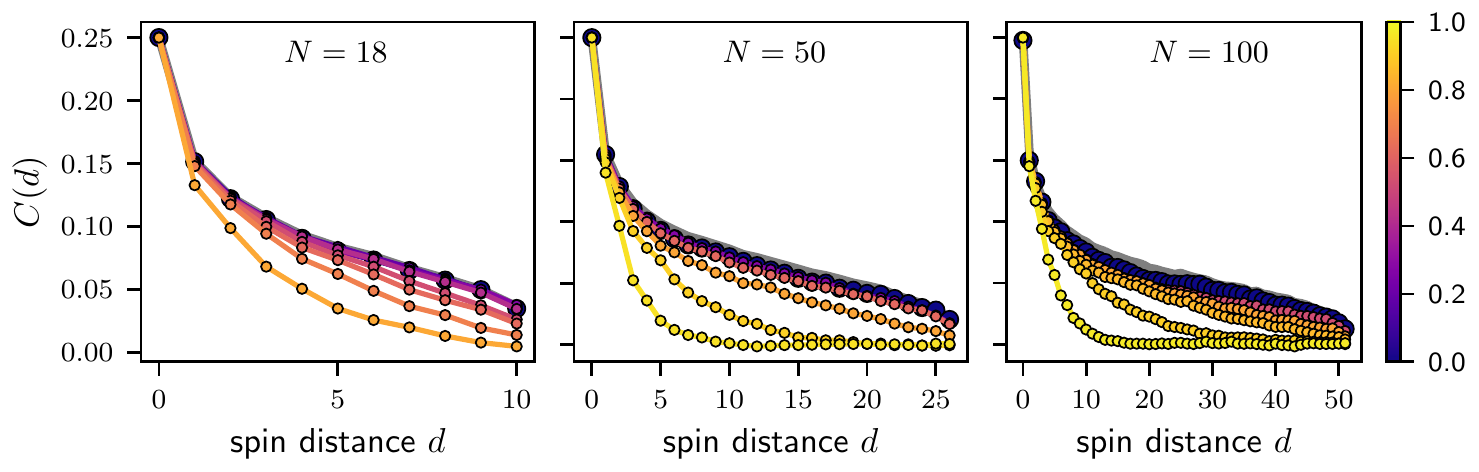}
    \vspace{-3mm}
    \caption{Spin-spin correlation function 
        versus the distance between lattice sites, as measured from the middle of the spin chain,
        for different system sizes (\textit{left:} 18; \textit{middle:} 50; \textit{right:} 100 spins).
        The color bar indicates the fraction of weights pruned. 
        The training set baseline is shown as a thick gray line, 
        which is mostly covered by the first plotted RBM-sample data (dark blue).}
    \label{fig:SSCF_vs_d}
\end{figure*}

The order parameter is the most pruning-sensitive observable.
At the QCP, mROE increases already at the first pruning iteration (albeit only weakly), 
indicating that pruning has an immediate effect on the order 
and the correlations between the spins.
We take a closer look at the order parameter, defined in Eq.~\eqref{eq:Magnet}, 
in Figure~\ref{fig:Magnet}:
The training set has $\langle |m| \rangle \approx 0.5$, 
indicating the presence of a preferred spin alignment direction \textit{within} configurations, 
and $\langle m \rangle \approx 0$, 
meaning that the two alignment directions are equally frequent. 
This is confirmed by the histogram of $m$ for the individual configurations (right panel of Figure~\ref{fig:Magnet}), which 
is symmetric around zero and shows a double-peak structure with maxima at $m=\pm 0.75$, 
but also a significant proportion at $m=0$.
We compute the occurrence ratio 
for a configuration $v$ and its $\mathbb{Z}_2$-symmetric counterpart $-v$ 
in the training set and find $0.98 \pm 0.19$, 
indicating that the training set is mostly $\mathbb{Z}_2$-symmetric, 
as mentioned in the previous section. 
The sample set drawn from the trained RBM 
resembles the properties of the training set closely. 
However, in the course of pruning, the double-peak structure morphs into 
an increasingly pronounced single peak at $m=0$. 
This change indicates that pruning induces loss of correlations between spins. 
In case when spins in the chain are completely independent, 
the two possible spin orientations, $+1$ and $-1$, become equally likely 
for every spin, 
yielding an expected value of zero for each $\sigma^z_i$ and consequently also for $m$. 
In Figure~\ref{fig:SSCF_vs_d} we observe this effect more directly: 
As a consequence of pruning, the functional form of the two-point correlation function 
changes from algebraic to exponential decay. 

These results align with our intuition that the physical system is most complex at the QCP, 
where the correlation length spans the entire system. 
We therefore expect that more parameters are required to model a system at a QCP. 
In contrast, correlations decay exponentially in the disordered PM phase, 
thus the model can be simpler and many weights in the RBM are superfluous.

Note the effect of system size in Figure~\ref{fig:SSCF_vs_d}: 
In larger models, a larger percentage of weights has to be pruned 
in order to induce the same degree of damage to the correlation function. 
The reason for why larger RBMs seem more robust 
is of combinatorial nature: 
The minimal number of edges required 
to keep a bipartite graph connected 
is one less than the total number of nodes. 
This means that at any given percentage of weights remaining, 
the graph of the smaller RBM is more likely to be disconnected 
than the graph of the larger RBM. 

\subsection{Training Sparse RBMs} \label{sec:results-Sparse}
Having observed that a trained RBM can be pruned to some extent without a significant loss in accuracy, 
we investigate whether an RBM that is sparse already \emph{at initialization} 
can be trained to achieve the same results as a dense RBM. 
In order to investigate this question in the context of RBMs applied to reconstructing our quantum wavefunction, 
we conduct two types of experiments with modified RBM weight matrices: 
In type 1, we apply a sparsity mask \textit{obtained by pruning} to an RBM  
with weights reset to 
(a) their original initialization values, or 
(b) other initial values (i.e., different random seed is used when drawing the initial weight values).
In type 2, we \textit{construct} the sparsity mask according to ad-hoc rules that we 
derive based on regularities that we observe in the weight matrix of trained RBMs (i.e.~its cluster structure). 
Namely, for our $N_h = 2N$ models, we observe that for each hidden unit, there is one cluster of three weights with significantly larger magnitude than the rest. 
The resulting rules are:
\begin{enumerate}
    \item Each hidden unit must be connected to a cluster of $X\geq 3$ adjacent visible units. 
    The value of $X$ is chosen according to the desired sparsity level. 
    \item The RBM graph must remain connected.
\end{enumerate}

We analyze the importance of the cluster structure for model performance and its relationship with the weight values at initialization in a series of ablation experiments. 
Our most important observations are the following.
Firstly, gradually breaking up the cluster structure of the pruning mask 
while maintaining the number of weights constant 
leads to a gradual decrease in model performance. 
Furthermore, when we prune the weights that would form clusters in the final model 
at initialization, 
the model can still be trained successfully 
and its final weight matrix will have clusters in other positions. 
Based on these observations we conclude: 
\begin{itemize}
    \item The cluster structure of the RBM weight matrix is crucial 
    for proper functionality of the generative model. 
    \item The position of the clusters is \emph{not} fully predetermined by the weights' initialization values.
\end{itemize}
The latter fact also aligns with the observation that 
the sparsity pattern obtained for an RBM initialized with some random seed 
works reasonably well for RBMs with other initialization seeds. 

\begin{figure*}[ht]
    \includegraphics[width=0.95\textwidth]{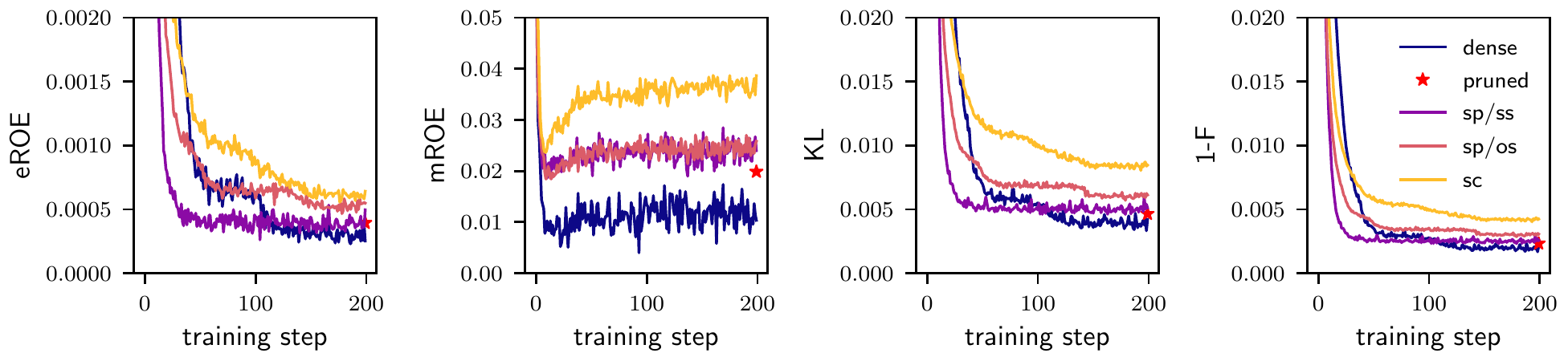}\\
    \includegraphics[width=0.95\textwidth]{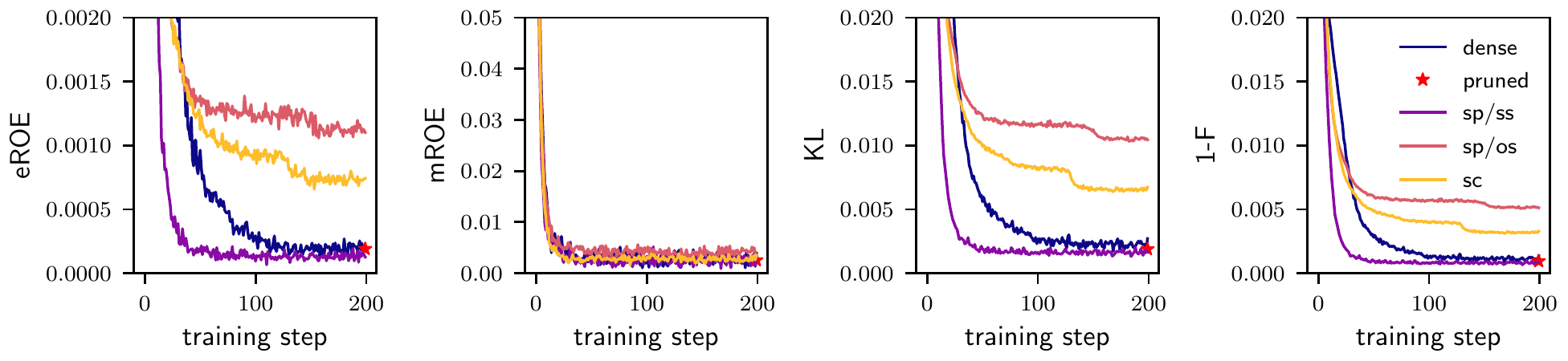}
    \caption{Training curves for various sparse RBMs 
    with approximately 60\% weights being nonzero and the dense RBM for comparison. Top row $h/J=1$, bottom row $h/J=2$.
    Legend abbreviations: \textbf{dense:} standard RBM, no weights removed; 
    \textbf{pruned:} trained dense and pruned post training for 5 iterations with fine-tuning;
    \textbf{sp:} trained sparse, sparsity pattern obtained by pruning of another model with same seed (\textbf{ss}) or other seed (\textbf{os});
    \textbf{sc:} trained sparse, sparsity pattern obtained by construction.}
    \label{fig:masked_RBM}
\end{figure*}

In Figure~\ref{fig:masked_RBM} we present the results of training sparse RBMs 
with different kinds of sparsity mask in comparison to dense RBM and pruned RBM. 
We consider both $h/J=1$ and 2, and choose a moderate value of sparsity -- 
the sparse RBMs have about 40\% weights removed. 
In each of the tested cases we find that a sparse RBM is trainable and 
the quality of the resulting model is comparable with the RBM that was trained dense and pruned post training. However, the differences can be substantial and do depend on the phase. 
In both regimes, the RBM with same-seed sparsity mask (sp/ss) 
shows faster convergence to its minimum errors. 
This fact is interesting from the optimization perspective -- 
intuitively, it indicates that removing unnecessary weights from the start 
allows the algorithm to traverse the optimization trajectory and reach the optimum faster. 
In the PM phase, the sp/ss RBM attains the smallest error values, 
outperforming even the dense RBM.
At the QCP, all sparse RBM variants do not outperform the pruned RBM, 
albeit the performance of the sp/ss RBM is close. 
A notable difference between the two regimes is that the worst-performing sparsity types are different:
In the PM phase, it is sp/os, while at the QCP it is the sc type.
This result indicates that the ad-hoc rules for constructing cluster-based sparsity masks 
might be missing some constraint for the QCP regime.

\section{Discussion}\label{sec:Discussion}

In this paper, we have examined the effects of pruning on a restricted Boltzmann machine (RBM)
trained to represent a quantum wavefunction 
based on simulated projective measurement data 
from the groundstate of the transverse-field Ising model in one dimension.  
We have observed that the effect of pruning varies drastically 
depending on the physical phase that the system is in. 
For RBMs trained to represent the groundstate in the paramagnetic (PM) phase, 
up to 50\% of weights can be pruned without significant loss of accuracy 
in the physical properties of the reconstructed state. 
In contrast, for RBMs trained on data at the quantum critical point (QCP), 
even a relatively small amount of pruning can have adverse 
effects on the model accuracy.
This result is intuitive, as at the QCP the state of the physical system is highly entangled and is characterized by long-range spin-spin correlations, 
while the PM phase the correlations decay rapidly. 
Therefore, we expect that the probability distribution that corresponds to the PM phase 
can be encoded in a simpler and less expressive function, and thus many weights in the dense RBM are superfluous.
An interesting question to pursue is whether this result has consequences 
beyond quantum critical systems, 
as much of the data from the natural world used to train neural networks in industry 
displays signatures of similar power-law decay~\cite{Ruderman_1994}.

Furthermore, our experiments demonstrate that pruning has disparate effects 
on various model quality measures. 
Specifically, we have found that among all physical observables energy is least sensitive to pruning.
More precisely, a pruned RBM can generate samples that have a reasonably accurate energy, 
while spin order and correlations show strong deviations from the ground truth.
Thus, for our learning task, 
energy should not serve as a measure for model quality when pruning is applied.
In general, 
our result demonstrates that it is important to monitor all relevant model quality measures 
instead of relying on a single observable. 
This is rather easy to accomplish when modeling a physical system, 
as the relevant physical observables are well-defined, 
but much harder for non-physical learning tasks, such as classification of natural images,
where apart from the empirical test error no other measure of model quality is readily available. 
The risk of pruning in such settings is that it might be introducing unnoticed instabilities into the model 
while keeping the standard test accuracy intact.
Such instabilities could then be exploited by adversaries or lead to biased outputs, 
undermining the reliability of the NN-based application.
Indeed, in ML context it is known that pruning significantly reduces robustness to 
image corruptions and adversarial attacks \footnote{An ``adversarial attack'' is 
an input image that was deliberately modified by introducing a small, 
typically unnoticeable variation, which causes a well-trained NN-based image classifier 
to make unreasonably wrong predictions.}~\citep{Xiao_2018,Guo_2019}.
Furthermore, it has been found that model performance is disproportionally impacted 
for classes of images that are generally more challenging to learn, 
which presents a concern for the fairness of AI algorithms~\cite{Hooker_2020, Paganini_2020}.

Finally, our study highlights the differences between pruning a fully-connected RBM before and after training. 
In the latter case, we observe that masks found by magnitude-based pruning of a dense RBM can be applied {\it a posteriori} to produce a sparser weight matrix; however, the accuracy of the resulting trained model can depend significantly on initial conditions for the weights. In the former case, we find that pruning masks defined through ad-hoc rules that take into account spatial locality and applied {\it a priori} often give poor accuracy. 
These observations suggest that pruning could be a viable strategy to reduce the computational resources
required by RBMs used in other contexts, for example in the variational setting~\cite{CarleoTroyer2017Science}.
However, significant further investigation would be required to devise pruning masks that could be applied 
to RBMs before training. 
This can be contrasted to Tensor Networks, 
where efficient wavefunction ansatze are constructed {\it a priori} based on assumptions of low entanglement~\cite{Chen_2018}. 
Thus, we hope that our study will motivate future investigations of pruning strategies for RBMs and other
generative models with applications both in quantum physics and to other natural phenomena.

\subsection*{Acknowledgments}
We acknowledge helpful discussions with E. Rrapaj and D. Sehayek.
The DMRG calculations were performed using ITensor~\cite{ITensor}.
RBM training was performed using the QuCumber package~\cite{qucumber}. 
This work was made possible by the facilities of the Shared Hierarchical Academic Research Computing Network (SHARCNET) and Compute Canada.
AG is supported by NSERC, Vector Institute for AI, BorealisAI, and NSF. RGM is supported by NSERC, the Canada Research Chair program, and the Perimeter Institute for Theoretical Physics. 
Research at Perimeter Institute is supported in part by the Government of Canada through the Department of Innovation, Science and Economic Development Canada and by the Province of Ontario through the Ministry of Economic Development, Job Creation and Trade.
This work is also supported by the National Science Foundation under Cooperative Agreement PHY-2019786 (The NSF AI Institute for Artificial Intelligence and Fundamental Interactions, http://iaifi.org/).
\bibliographystyle{apsrev4-1}
\bibliography{bibliography.bib}

\newpage

\section*{Appendix}

\subsection{Pruning schedule}\label{appx:schedule}

In this study, we chose to prune 10\% of the weights in the RBM at every iteration. In Table~\ref{tab:T1} we provide a list of the pruning iterations and the respective percentage of weights remaining in the RBM. 
We have tested alternative pruning schedules, where we prune a larger percentage of weights in the first iteration (between 10\% and 40\%) and a smaller percentage in the following iterations (5\%). Experiments with these schedules have not revealed any qualitative difference.

\begin{table}[h]
    \begin{tabular}{c|c}
    pruning iteration & weights remaining \\
    \hline \hline
        0 & 100.0 \% \\
        1 & 90.0 \% \\
        2 & 81.0 \% \\
        3 & 72.9 \% \\
        4 & 65.6 \% \\
        5 & 59.0 \% \\
        6 & 53.1 \% \\
        7 & 47.8 \% \\
        8 & 43.0 \% \\
        9 & 38.7 \% \\
        10 & 34.9 \% \\
        11 & 31.4 \% \\
        12 & 28.2 \% \\
        13 & 25.4 \% \\
        14 & 22.9 \% \\
        15 & 20.6 \% \\
        16 & 18.5 \% \\
        17 & 16.7 \% \\
        18 & 15.0 \% \\
    \end{tabular}
    \caption{Percentage weights left in the RBM at a given pruning iteration. 
    The RBM graph becomes necessarily disconnected 
    at the 18th pruning iteration 
    when less than 26 weights (16.05\%) are left.}
    \label{tab:T1}
\end{table}

\subsection{Comparison of error measures}\label{appx:ROE_KL_correlat}

\begin{figure}[h]
\includegraphics[width=0.75\columnwidth]{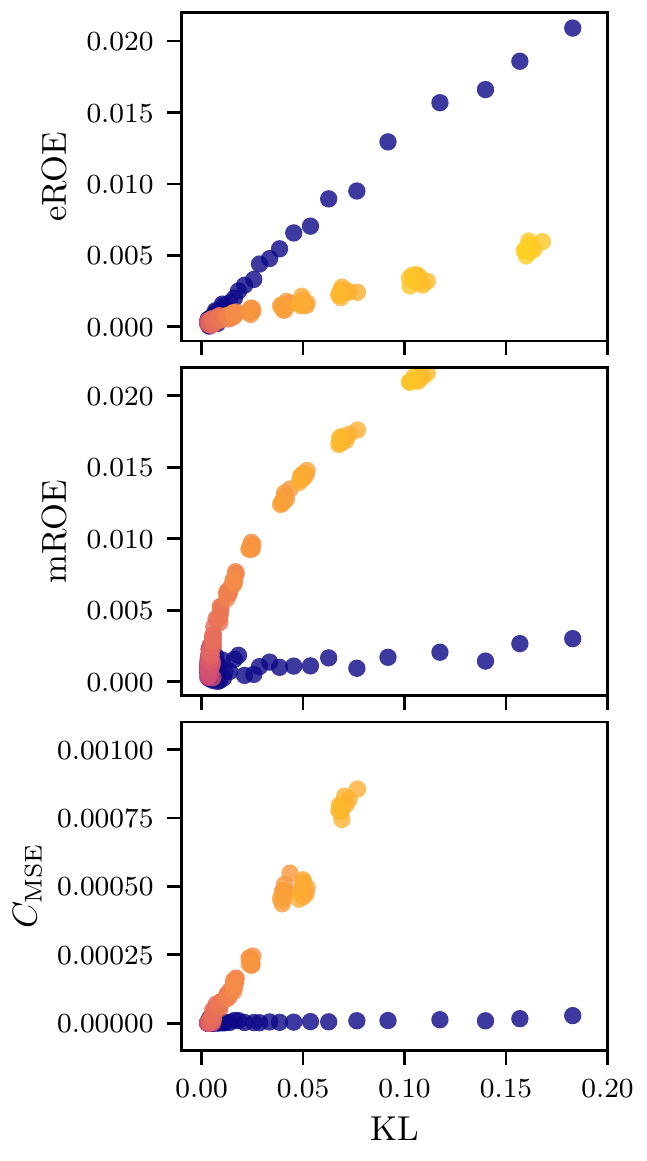}
\caption{Correlation between physical error measures and the KL divergence 
    during training and pruning, up to the 18th pruning iteration (where the RBM graph necessarily becomes disconnected). 
    Training phase is indicated by blue markers, pruning phase by a color gradient from purple to yellow.}
\label{fig:ROEs_vs_KL}
\end{figure}

In this study we have defined a number of 
model quality measures based on physical observables that 
we have used additionally to the standard KL divergence. 
One reason for introducing these error measures is that 
here we are concerned with learning a model for a physical system; 
it is therefore an essential requirement to ensure that 
the physical observables are correct.
Another reason is that the computation of the KL divergence is not feasible for large systems, 
and thus we have to rely on other quality measures, as discussed in Section~\ref{sec:Results}. 
While the CD training procedure by construction minimizes the KL divergence, 
the physical observables are not optimized for. 
In Figure~\ref{fig:ROEs_vs_KL} we present evidence that our physical measures 
are indeed correlated with the KL divergence. 
Note, however, the distinct behavior: 
mROE and the MSE of the 2-point correlation function $C_\text{MSE}$ 
are significantly more sensitive to pruning than KL, while eROE is less sensitive. 
These differences demonstrate the necessity of tracking all relevant physical observables 
when pruning an RBM to ensure that the model remains correct.

\end{document}